\documentclass[%
 reprint,
 amsmath,amssymb,
 aps,
]{revtex4-2}

\usepackage{graphicx}
\usepackage{dcolumn}
\usepackage{bm}
\usepackage[version=3]{mhchem} 

\usepackage[usenames]{color}
\usepackage[normalem]{ulem}

\begin{document}

\preprint{APS/123-QED}

\title{\textbf{Transformation of the gyrotropic mode spectrum of the FM / AFM disk via vortex imprinting} 
}%
\author{Evgeny V. Skorokhodov$^{1}$}
 \email{evgeny@ipmras.ru}
\author{Evgeny A. Karashtin$^{1,2,3}$}
 \email{eugenk@ipmras.ru}
\author{Ilya A. Fedotov$^{1}$}
\author{Igor Yu. Pashen'kin$^{1}$}%
\author{Maksim V. Sapozhnikov$^{1,2,3}$}

\affiliation{$^{1}$Institute for Physics of Microstructures RAS, Nizhny Novgorod,  Russia}
\affiliation{$^{2}$Lobachevsky State University of Nizhny Novgorod, Nizhny Novgorod, Russia}
\affiliation{$^{3}$ MIREA – Russian Technological University, Moscow, Russia}%

\date{\today}

\begin{abstract}
We experimentally investigate the magnetic gyrotropic mode in a system of vortex ferromagnetic (FM) nanooscillator exchange coupled to an antiferromagnetic (AFM) layer. The micron-sized disks formed from the \ce{Ni80Fe20}(12 nm) / \ce{Ir80Mn20} (5 nm) FM/AFM heterostructure are prepared so that the vortex magnetic state is imprinted into the AFM layer. We apply a magnetic resonance force microscopy (MRFM) method to locally study magnetic oscillations in single FM/AFM disks. We show that the gyrotropic mode frequency is significantly (approximately four times) shifted to the high frequency range compared to a similar structure consisting of a single ferromagnetic disk. Upon applying an in-plane magnetic field, we observe a strong peak at a double frequency which was previously predicted in theory. This shows that nonlinearity in vortex dynamics is strongly enhanced in the FM/AFM system under investigation. Thus the magnetic imprinting technology reveals great potential for future application in high-frequency spintronic devices.  
\end{abstract}

\maketitle


In modern science, a lot of effort is devoted to the creation, investigation, and control of nonuniform magnetic states. This is governed by great potential for application of such systems: information storage, magnetic field detection, electromagnetic wave generation, etc \cite{Tulapurkar2005, Chappert2007, Kiselev2005, Yuasa2004, Parkin2004}. Such devices are scalable from macro- to nanosizes. The most simple way to create such nonuniform states is to make patterned magnetic nanostructures. In this case the magnetostatic interaction does all the work, overcoming the exchange and crystalline anisotropy. A highly symmetric and simple example of such nonuniform magnetic structure stabilized by magnetostatic mechanism is the vortex. It contains inexaustible possibilities and may be used for all the mentioned applications. Vortex magnetization distribution is typically realized in nanodisks with more than 100~nm diameter and thickness greater than 10~nm \cite{Usov1993, Shinjo2000, Gregurec2020, Gonzalez2024}.

On the other hand, nonuniform antiferromagnetic order is much less studied. Although there are certain suggestions of nonuniform antiferromagnet based devices there are certain problems on the way to implement antiferromagnetic spintronics. Another problem is that the antiferromagnet has no net magnetization, so detection and visualization of nonuniform antiferromagnetic state is usually a challenge \cite{Nolting2000, Evans2002}. Besides, there are no magnetostatic fields due to locally compensated magnetic moments which leads to a problem of stabilizing the nonuniform states in antiferromagnets. Quite often they are pinned to local structure defects \cite{Kleemann1993} which makes it hard to exploit such states. 

One property of antiferromagnets that is widely used is their ability to shift the magnetization loop of adjacent ferromagnets due to the surface exchange interaction. This phenomenon is important for lots of spintronic devices such as magnetic memory or field sensors. The exchange coupling between the ferromagnet and the antiferromagnet in such systems may be used to obtain indirect data regarding the magnetic state of the antiferromagnet, e.g. the spin-flop field, the domain wall width, or the antiferromagnetic order at surface. Moreover, the ferromagnet may influence the state of the antiferromagnet via this exchange coupling \cite{Bruck2005, Petracic2005, Roshchin2005}. Particularly, a nonuniform vortex state was shown to be imprinted from a ferromagnet (FM) to an antiferromagnet (AFM) in the micron-sized FM/AFM bilayer disks \cite{Sort2006}. In order to obtain this, the sample is heated to a temperature greater than the blocking temperature of the AFM in a zero magnetic field. This leads to a formation of vortex state in the FM disk which is subsequently set in the AFM disk at cooling down to room temperature. The exchange coupling between the layers is responsible for this possibility, and the same exchange coupling leads to the magnetization loop change for the FM layer: The vortex state becomes more stable in the vicinity of zero external magnetic field.

Structured magnetic systems with vortex magnetization distribution are the basic element of vortex spin-transfer nanooscillators (VSTNO). The VSTNO systems may act as the sources or the detectors of electromagnetic waves \cite{Skirdkov2018, Zvezdin2022}. The VSTNO is a magnetoresistive contact (usually tunnel magnetic junction) in which the free layer has vortex magnetization. The working principle of these devices is based on two cornerstones: First, the vortex magnetic moment distribution posesses a gyrotropic oscillation mode (with typical frequency of 0.1-1~GHz) \cite{Thiele1973, Ivanov2007, Gaididei2010, Iurchuk2024}, and second, this mode may be excited by a constant spin-polarized electric current in a self-oscillation mode. Certain ways to control the gyromode frequency in the range of 0.1-1~GHz are based on its dependence on the disk geometry and size, the magnitude of the spin-polarized electric current, or on the applied external magnetic field. All these methods have a strong limitation connected to the destruction of proper vortex distribution due to change of the particle size, or in a high current or magnetic field. Another way to modify the gyration mode in a vortex is to pin it by a defect \cite{Compton2006}; however it is quite hard to manipulate the properties of such system by external influence. A similar yet more efficient possibility to control the VSTNO frequency is to use a magnetic vortex coupled to an AFM layer with imprinted vortex state. Theoretical estimations show an order-of-value rise in frequency which may be controlled simply by changing the FM layer thickness. This brings the VSTNO into the gigahertz frequency range which may broaden their usage in applications \cite{Buchanan2008, Heinonen2007}.

This paper is devoted to an experimental study of gyrotropic modes in micron-sized disks made of a Ta(20~nm)~/ IrMn(5~nm)~/ Fe(0.5~nm)~/ NiFe(12~nm)~/ Ta(5~nm) film made on a thin glass substrate. We show a successful imprinting of vortex state into the IrMn layer and use the magnetic resonance force microscopy (MRFM) method to detect the oscillations of a single disk. A comparison to similar disk made of Ta(20~nm)~/ NiFe(12~nm)~/ Ta(5~nm) film shows a significant (approximately four times) growth of the gyromode frequency. Besides, we evidently show that the double-frequency mode is strongly excited in an applied constant magnetic field due to the highly noncentrosymmetric structure of the sample.

\begin{figure}[tb]
\centering{\includegraphics[width=0.7 \hsize]{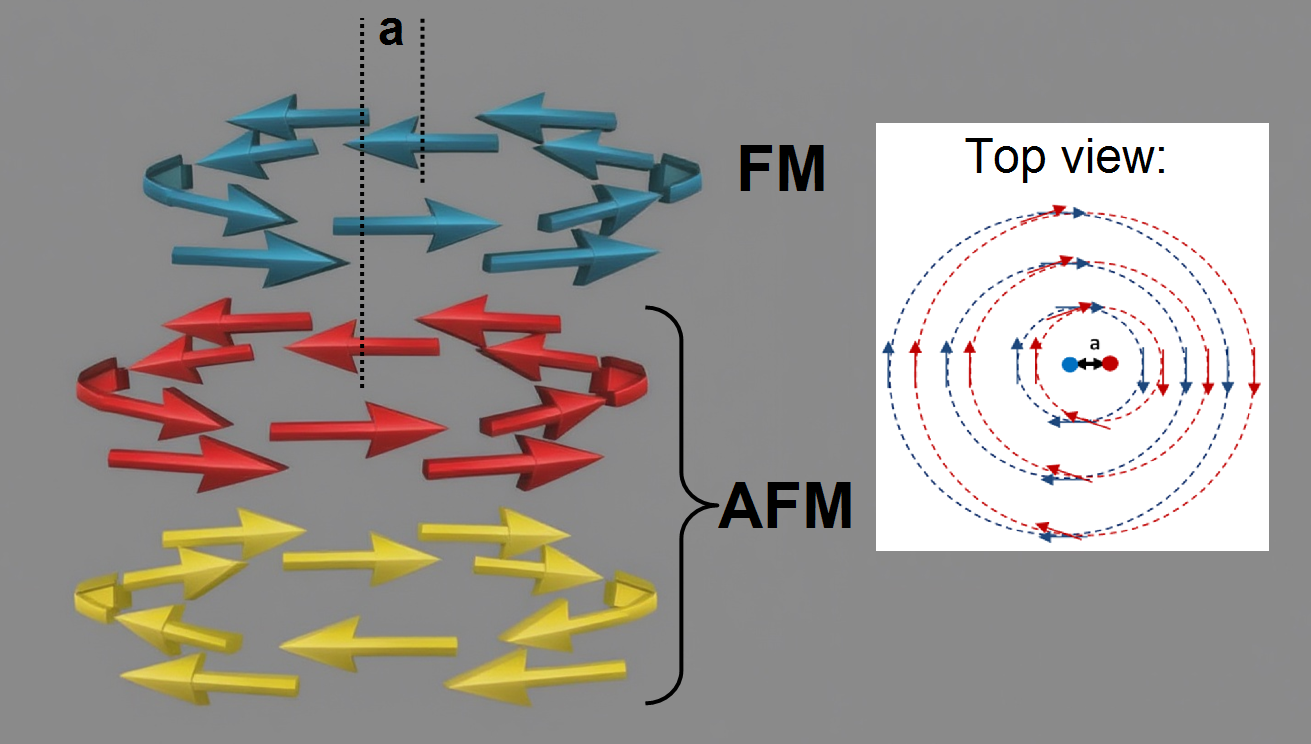}}
\caption{\label{fig4} The illustration of the exchange energy growth due to the FM vortex shift with respect to the AFM vortex.}
\end{figure}

The surface exchange interaction between vortices in ferromagnet and antiferromagnet leads to the resonance frequency shift. Figure~\ref{fig4} schematically shows the mechanism of the influence of the exchange coupling between the FM and the AFM on the gyromode frequency. The shift of magnetic vortex from the disk center leads to the exchange energy growth. This addition to energy leads to the growth of gyromode frequency which can be expressed as
\begin{equation} \label{Eq1}
    \omega_{exch} \approx \gamma H_{exch} \ln{\frac{R}{d}}
\end{equation}
in the approximation of small core shift $a$ with respect to the disk radius $R$. Here $d$ is the core radius, $H_{exch}$ is the exchange field that acts on the ferromagnet and is induced by the antiferromagnetic layer (note that $H_{exch}$ is inversely proportional to the ferromagnetic layer thickness).

\begin{figure}[tb]
\centering{\includegraphics[width=0.7 \hsize]{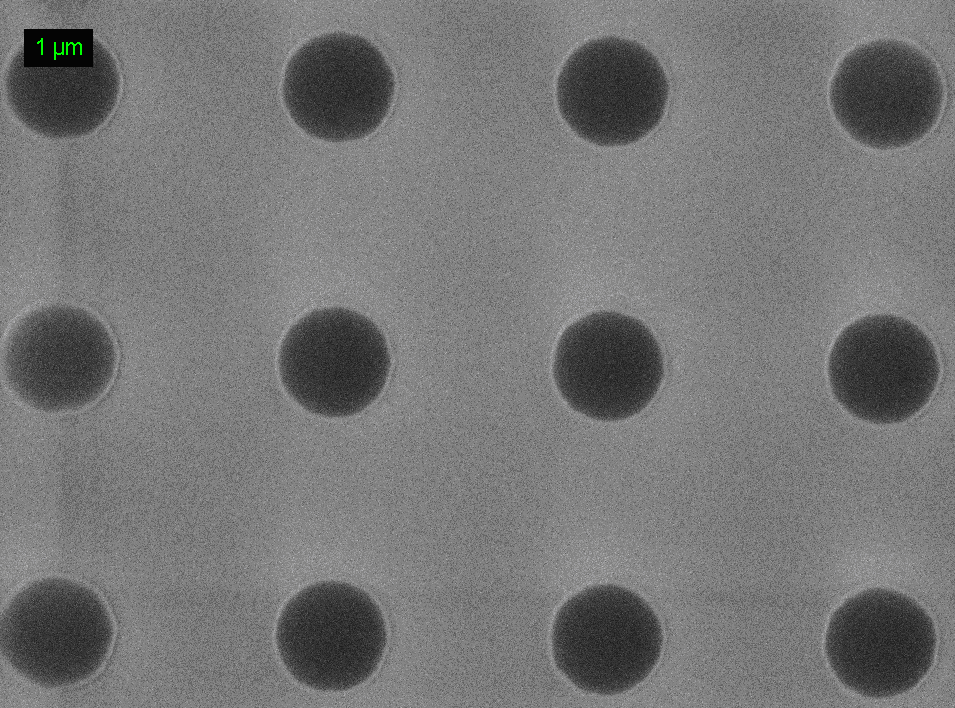}}
\caption{\label{fig1} Scanning electron microscope (SEM) image of a fragment of the disk array. The disk diameter and the distance between edges are $1~\mu m \pm 10nm$.}
\end{figure}

\begin{figure}[tb]
\centering{\includegraphics[width=\hsize]{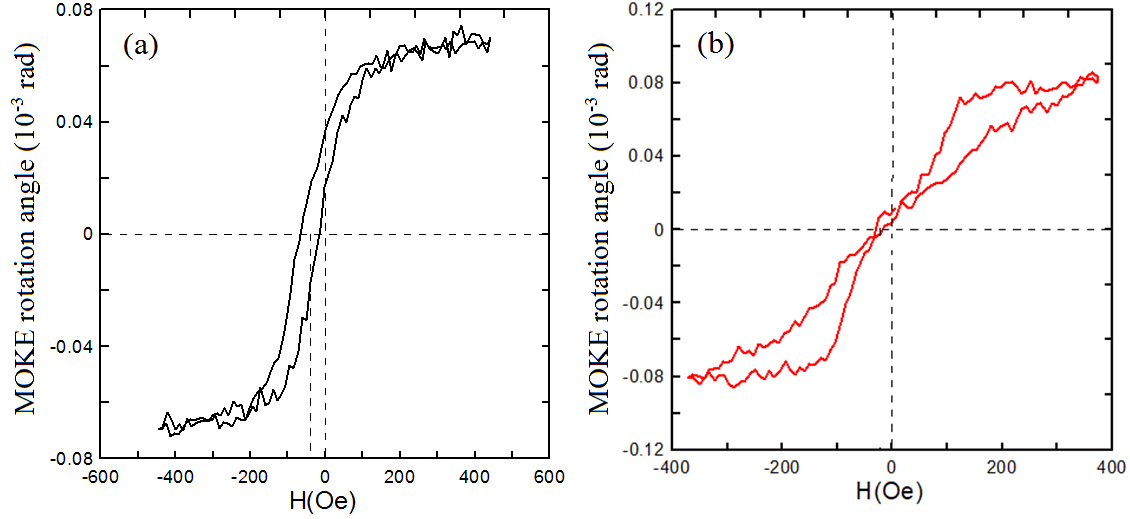}}
\caption{\label{fig2} Magnetization curves of an array of disks made of a Ta(20~nm)/IrMn(5~nm)/Fe(0.5~nm)/NiFe(12~nm)/Ta(5~nm) film (a) before annealing and (b) after annealing at $350^\circ C$ and subsequent zero-field cooling.}
\end{figure}

A Ta(20~nm)~/ IrMn(5~nm)~/ Fe(0.5~nm)~/ NiFe(12~nm)~/ Ta(5~nm) film is obtained by magnetron sputtering. The residual pressure in the sputtering chamber does not exceed $5 \cdot 10^{-7} Torr$, and the working pressure of argon during the sputtering process is $2 \cdot 10^{-3} Torr$. The layers are in the order from substrate to surface. After that, a $300 \times 300 \mu m^2$ array of micron disks with a micron distance between the edges is formed by electron lithography and ion etching (Figure~\ref{fig1}).
The size of the disk array is chosen in order to characterize their net state by the magnetooptical Kerr effect (MOKE) method. Figure~\ref{fig2}~(a) contains a longitudinal MOKE loop of the array measured directly after fabrication. The magnetic field is applied along the easy axis of the structure inherited from the film. One can see that there is an exchange bias of the order of $100Oe$. 

In order to obtain an imprinting of vortex state into the antiferromagnetic IrMn, the structures are thermally annealed at a temperature of $350^\circ C$ which exceeds the IrMn blocking temperature. After that, they are cooled in a zero field. The magnetization curve of disk arrays after annealing and zero-field cooling is presented in Figure~\ref{fig2}~(b). We see a hysteresis-less behavior of the magnetic moment close to zero field which corresponds to the vortex state, in a similar way to literature \cite{Sort2006}. Thus, vortex state imprinting into antiferromagnet is realized in the fabricated disk array. 

\begin{figure}[tb]
\centering{\includegraphics[width=0.8\hsize]{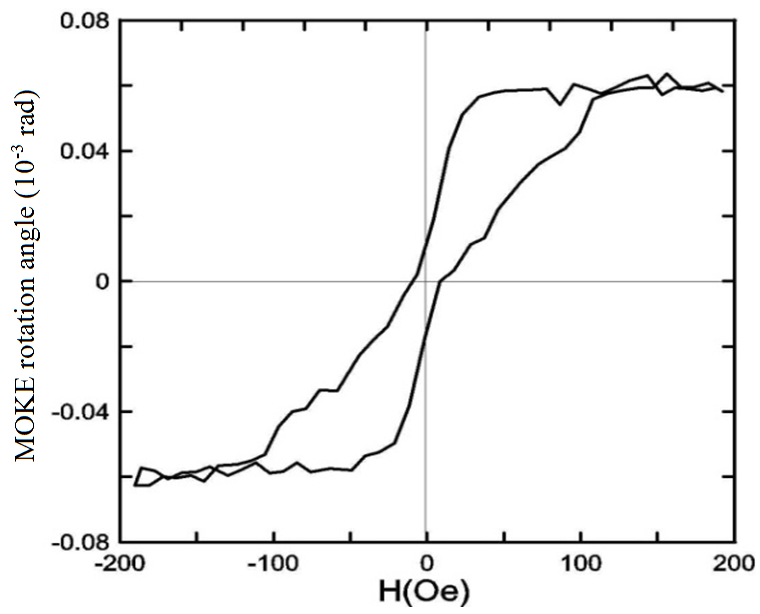}}
\caption{\label{fig3} MOKE magnetization curve of a micron-sized disk array made of a Fe(0.5~nm)/NiFe(12~nm)/Ta(5~nm) film.}
\end{figure}

We also create a reference sample of similar disk array made of a  Fe(0.5~nm)/NiFe(12~nm)/Ta(5~nm) ferromagnetic film without an antiferromagnetic layer. Magneto-optical studies show that there are vortex magnetic states in the obtained disks confirmed by a close to zero residual magnetization (Figure~\ref{fig3}). However there is still hysteresis which is explained by less stable vortex state and inhomogeneities from disk to disk.

Magnetic resonance force microscopy (MRFM) method is used to study the resonant properties of individual disks \cite{Lopes2025} in which vortex imprinting into an antiferromagnet is implemented. The MRFM microscope is a scanning probe microscope equipped with a microwave pumping system. The dependence of the amplitude of the probe oscillations on the frequency of microwave pumping is recorded. To increase the sensitivity of the measurements, the microwave field is modulated at the resonant frequency of the cantilever. Cantilevers with a magnetic coating (CoFe film about 300 nm thick) and low stiffness (~ 0.1-0.03 N/m) are used as probes. All measurements are carried out in a vacuum in order to increase the Q-factor of the cantilever (Q~2000). A more detailed description of MRFM may be found elsewhere \cite{Skorokhodov2021, Skorokhodov2023}.

\begin{figure}[tb]
\centering{\includegraphics[width=0.8 \hsize]{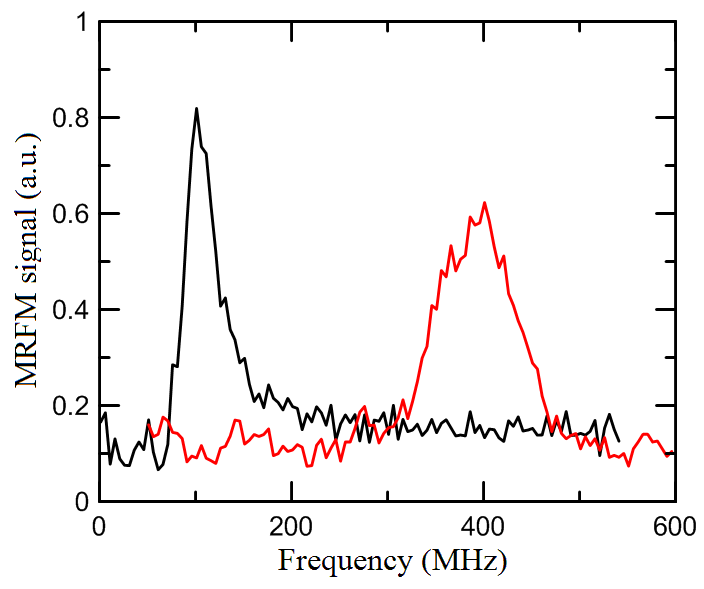}}
\caption{\label{fig5} MRFM spectra for a single FM disk (black line) and an FM/AFM disk with a vortex imprinted into the AFM layer (red line).}
\end{figure}

Figure~\ref{fig5} shows the ferromagnetic resonance spectra obtained by magnetic resonance force microscopy for a ferromagnetic disk and for a disk made of an FM/AFM film. One can see that the resonance frequency of the gyrotropic mode for a bilayer disk made shifts significantly (approximately 4 times) to the high frequency range.

\begin{figure} [tb]
\centering{\includegraphics[width=\hsize]{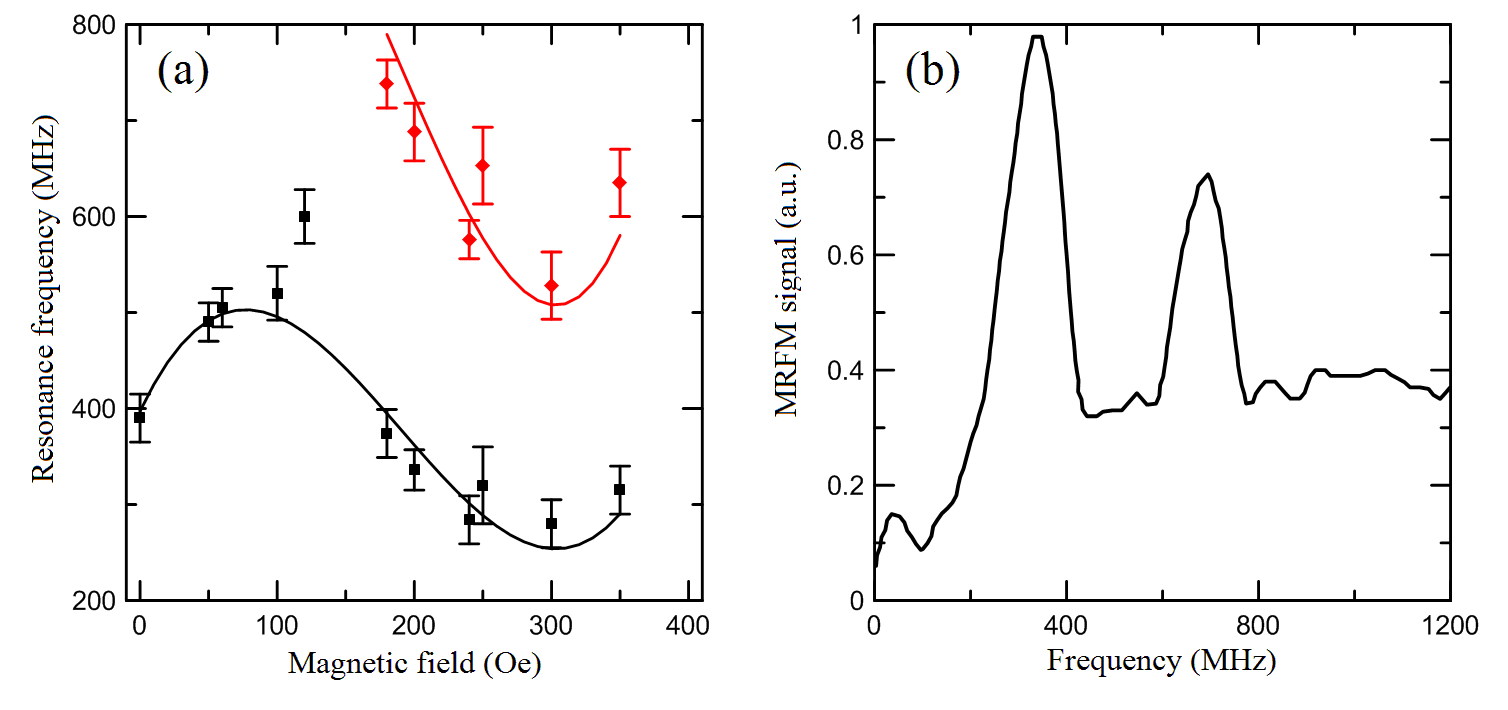}}
\caption{\label{fig6} (a) Dependence of the observed resonance frequencies of the gyrotropic oscillations of the FM vortex with respect to the applied magnetic field. Red line is doubled black line. (b) MRFM spectrum for the magnetic field $H=200~Oe$. Two peaks are clearly observed.}
\end{figure}

Next, we measure the dependence of the gyromode on a constant external magnetic field applied in the disk plane. In theory, the frequency should decrease once the FM vortex is sufficiently shifted from an energy minimum created by the AFM layer. In experiment, we observe a slight growth of the frequency for the field up to $120~Oe$, which seem to correspond to a small vortex shift, and a significant decrease for greater field (Figure~\ref{fig6}~(a)). This may be described by a concurrence between the Zeeman energy and the boundary exchange coupling to the AFM layer which are both strongly nonlinear in $H$. The increase in Zeeman energy dominates at small fields which leads to a frequency growth. For higher fields, the FM vortex shifts from the AFM vortex significantly and thus oscillates in an almost uniform exchange field which leads to a significant diminishing of the correction (\ref{Eq1}) to the gyromode frequency. At high field, the frequency tends to that of a single FM layer ($300~MHz$ for $H=300~Oe$), which shows that the AFM correction almost vanishes. A slight growth of the frequency is explained by the Zeeman energy, once again. The vortex passes out of the disk at $H = 400~Oe$ (Figure~\ref{fig2}~(b)); we do not register the known decrease of the gyromode frequency at high field which is explained by the growth of the exchange energy inside the FM layer.

An additional peak is clearly observed for the field greater than $160~Oe$ (Figure~\ref{fig6}~(b)). The frequency dependence comparison of the high-frequency peak (red line in Figure~\ref{fig6}~(a)) and the low-frequency peak shows that the frequency ratio of these two peaks is 2 within the estimated error interval. The black line in Figure~\ref{fig6}~(a) approximates the first peak frequency dependence on field. The red line is the black line multiplied by 2; it approximates the second peak frequency dependence quite well. So we experimentally observe a double-frequency gyromode resonance which was theoretically studied in literature \cite{Lee2007, Guslienko2010}. Note that the MRFM experimental setup registers the $z$-component of the magnetization oscillations \cite{Skorokhodov2021, Skorokhodov2023} which leads to a strong signal from the double-frequency mode.

The system of exchange-coupled vortices in the FM and AFM layers possesses the break of inversion symmetry which leads to significant nonlinear terms in the potential energy $W$ \cite{Wang2026}. In the absence of the external magnetic field, $W$ contains only even in the core shift terms:
\begin{equation} \label{Eq2}
W = W_0 + k_2 a^2 + k_4 a^4 + ...
\end{equation}
Application of an external magnetic field $H$ leads to an additional in-plane asymmetry of the system which results in the odd in the core shift terms in energy:
\begin{equation} \label{Eq3}
W = W_0 + k_1 H a + k_2 a^2 + k_3 H a^3 + k_4 a^4 + ...
\end{equation}
These odd terms in $a$ lead to a double-frequency mode\cite{Lee2007, Guslienko2010}. Thus, a strong nonlinearity in the FM magnetization oscillations induced by a neighboring AFM layer which breaks the inversion symmetry of the system enhances the double-frequency oscillations and makes it possible to experimentally observe such a mode.

The obtained result is confirmed by micromagnetic calculations. We use the MuMax3 package with the $256 \times 256 \times 14$ grid. The grid cell size is $3.9 \times 3.9 \times 1.25 nm^3$. The AFM layer is modeled as a series of four successive layers (total thickness is $5~nm$) with opposite vortex magnetization and a very big anisotropy constant; the exchange between the FM and the neighboring AFM layer is modified in order to obtain the magnetization loop close to that observed experimentally (Figure~\ref{fig2}~(b)). The magnetic resonance spectra are calculated under the action of a small alternating magnetic field of $1~Oe$. We see from our calculation that if a magnetic field is applied in the layers plane second harmonic gyromode appears in the ferromagnetic disk (see Supplementary Material, Figure~S2).

To conclude, we experimentally show that the gyromode frequency of a micron-sized ferromagnetic disk can be effectively controlled by the magnetic state of a neighboring antiferromagnetic layer. The exchange coupling between two magnets leads to imprinting of a vortex state into the antiferromagnet upon heating above the blocking temperature and subsequent zero field cooling. The same coupling stabilizes the vortex state in a ferromagnet at room temperature and thus sufficiently increases the gyromode frequency: we observe an approximately fourfold increase for the IrMn(5~nm)/NiFe(12~nm) material combination at zero field. Application of an external field leads to a nonlinear behavior of frequency: a slight (approximately 1.5 times with respect to zero-field frequency) increase of the gyromode frequency at small field is followed by a pronounced decrease at higher field. 
Thus, the magnetic imprinting technology in a nanolayered system shows up a great potential of the controllable magnetic oscillations frequency manipulation.

Besides, nonlinear magnetization dynamics is strongly enhanced in via the imprinting. This  allows us to observe a significant double-frequency peak in an applied constant external field. This is supported by micromagnetic simulations and may be explained by the inversion symmetry break due to the boundary exchange coupling to the antiferromagnet. The imprinting technology in FM/AFM systems demonstrates high potential that may be subsequently used for future THz applications due to a significant blue shift of the eigenmodes. Although we achieve the frequency of less than $1~GHz$ for the gyration mode there are two possibilities: frequency enhancement of different modes such as uniform precession and generation of high harmonics due to strong nonlinearity of the system. Our results open up new frontiers in the construction of flexible vortex nanooscillators with a broadened spectrum range and strong nonlinear behavior for new spintronic devices.


This work is supported by the Russian Science Foundation, Grant No.~25-22-00670.

\bibliography{mrfm_fm_afm}

\end{document}